\documentclass[sigconf]{acmart}
\AtBeginDocument{%
  }

\copyrightyear{2026}
\acmYear{2026}
\setcopyright{cc}
\setcctype{by}
\acmConference[CHI EA '26]{Extended Abstracts of the 2026 CHI Conference on Human Factors in Computing Systems}{April 13--17, 2026}{Barcelona, Spain}
\acmBooktitle{Extended Abstracts of the 2026 CHI Conference on Human Factors in Computing Systems (CHI EA '26), April 13--17, 2026, Barcelona, Spain}
\acmDOI{10.1145/3772363.3798382}
\acmISBN{979-8-4007-2281-3/2026/04}

\usepackage{makecell}
\usepackage{enumitem}

\begin{document}

\title{``Oops! ChatGPT is Temporarily Unavailable!'': A Diary Study on Knowledge Workers’ Experiences of LLM Withdrawal}

\author{Eunseo Oh}
\authornotemark[1]
\affiliation{%
  \institution{KAIST}
  \city{Daejeon}
  \country{Republic of Korea}
}
\email{so.eso@kaist.ac.kr}

\author{Suyoun Lee}
\authornotemark[1]
\affiliation{%
  \institution{KAIST}
  \city{Daejeon}
  \country{Korea}}
\email{jenslee705@kaist.ac.kr}

\author{Jae Young Choi}
\authornote{All three authors contributed equally to this research.}
\affiliation{%
  \institution{KAIST}
  \city{Daejeon}
  \country{Korea}
}
\email{jaeyoungchoi@kaist.ac.kr}

\author{Soobin Park}
\affiliation{%
\institution{KAIST}
  \city{Daejeon}
  \country{Republic of Korea}}
\email{soobinpark@kaist.ac.kr}

\author{Youn-kyung Lim}
\authornote{Corresponding author.}
\affiliation{%
  \institution{KAIST}
  \city{Daejeon}
  \country{Korea}}
\email{younlim@kaist.ac.kr}

\renewcommand{\shortauthors}{Oh et al.}

\begin{abstract}
LLMs have become deeply embedded in knowledge work, raising concerns about growing dependency and the potential undermining of human skills. To investigate the pervasiveness of LLMs in work practices, we conducted a four-day diary study with frequent LLM users (N=10), observing how knowledge workers responded to a temporary withdrawal of LLMs. Our findings show how LLM withdrawal disrupted participants’ workflows by identifying gaps in task execution, how self-directed work led participants to reclaim professional values, and how everyday practices revealed the extent to which LLM use had become inescapably normative. Conceptualizing LLMs as infrastructural to contemporary knowledge work, this research contributes empirical insights into the often invisible role of LLMs and proposes value-driven appropriation as an approach to supporting professional values in the current LLM-pervasive work environment.
\end{abstract}




\maketitle

\section{Introduction}
Large Language Models (LLMs) have rapidly become an essential layer of everyday work. 
Within three years of its launch in November 2022, ChatGPT has reached hundreds of millions of weekly active users worldwide and accounts for billions of queries each week~\cite{chatterji2025people, acemoglu2025simple}. Microsoft's 2024 labor-market survey likewise suggests that three-quarters of knowledge workers already use LLMs at work~\cite{microsoft2024worktrend}, indicating that LLMs are becoming integrated in our daily work through which people write, code, search, and make sense of information.
While prior works document substantial productivity gains from LLM assistance in knowledge work~\cite{noy2023experimental, brynjolfsson2025generative}, researchers also warn that LLM use requires caution due to risks such as hallucination, privacy leakage, and harms regarding fairness and inequality ~\cite{teubner2023welcome, ji2023survey, weidinger2022taxonomy, bender2021dangers}.
In particular, concerns are growing around LLM dependency—over-reliance that may erode learning or skill development in education~\cite{kasneci2023chatgpt} and potentially contribute to knowledge collapse in creative works~\cite{peterson2025ai}. 
Recent studies also suggest that heavier reliance on LLMs can be associated with reduced cognitive effort and confidence in one’s own judgment among knowledge workers, as well as shifts in student engagement and academic achievement pathways mediated by self-efficacy~\cite{lee2025impact, lo2024influence, jia2025effect}.
Reflecting this emerging concern, recent works have begun to characterize and measure LLM dependency via instruments such as LLM-D12~\cite{yankouskaya2025llm} and the LLM Dependence Scale~\cite{li2025assistance}. 

However, most existing studies examine LLM use while the technology is present and available, leaving limited empirical insight into how deeply these services are embedded in everyday practice that can only be revealed when assistance is abruptly removed. 
As Star argued, it is often through breakdowns or outages that the true nature and scale of a technology become visible to users~\cite{star1999ethnography}. This dynamic became especially salient in November 2025, when a major Cloudflare outage disrupted global services, including ChatGPT~\cite{guardianCloudflare2025}. News coverage of the resulting disruption to knowledge work highlighted how strongly workers had come to rely on LLMs for daily tasks~\cite{dongaChatGPT2025, guardianCloudflare2025}.
In academia, prior studies on smartphone and social media abstinence similarly show that withdrawal help participants surface patterns of reliance and reflect on their relationship with technology \cite{WILCOCKSON2019106013,rosenberg2025missing}.
Therefore, we address the research question: \textit{\textbf{“What does the withdrawal of LLMs reveal about its pervasiveness in work?”}}
We conducted a diary study with knowledge workers\footnote{We use ``knowledge workers'' to refer to both employed knowledge workers and university students. Following Drucker’s definitions that characterize knowledge workers by the premise of developing or using knowledge and information in their main work \cite{drucker1959landmarks}, we include university students because their major day-to-day work activities substantially overlap with knowledge-work activities. } during a four-day period of LLM withdrawal, followed by semi-structured interviews.
Through this approach, we identify gaps in work practices that emerge during withdrawal and illustrate how workers’ perceptions of LLMs are reshaped through the experience. We further derive insights for fostering more reflective and intentional adoption of LLMs in work practices.


\section{User Study}

\textbf{\textit{Participants Recruitment.}} 11 participants were recruited through online platforms in South Korea.
We screened participants based on frequent use of and dependency on LLM services, measured utilizing LLM-D12 scale~\cite{yankouskaya2025llm}, while ensuring diversity in occupational backgrounds. We used only the \textit{instrumental dependency} items from the LLM-D12, as \textit{relational dependency} was out of scope.
Detailed demographic information is presented in Table~\ref{demo}. 
One participant was excluded from the analysis due to non-adherence to the study protocol,
our final dataset consisted of 10 participants (6 female, 4 male).
Each participant received a compensation of 70{,}000 KRW ($\approx\$48$) upon completion of study.

\textbf{\textit{Study Procedure.}} The study consisted of (i) a four-day LLM withdrawal diary study and (ii) a semi-structured interview, an approach commonly used to study technology non-use and deprivation ~\cite{almoallim2023patterns,rosenberg2022mobile,rosenberg2025missing,baumer2018departing,hall2021experimentally}. 
We decided on the period of four consecutive working days to capture work disruptions caused by LLM absence while considering participant burden.
During the diary study period, participants were asked to voluntarily refrain from using LLM-based services. 
If participants found voluntary refraining difficult and expressed a need for additional support, they were guided to install Chrome extensions that block access to LLM services\footnote{\href{https://chromewebstore.google.com/detail/blocksite-block-websites/eiimnmioipafcokbfikbljfdeojpcgbh}{BlockSite} and
 \href{https://chromewebstore.google.com/detail/bye-bye-google-ai-turn-of/imllolhfajlbkpheaapjocclpppchggc}{Bye Bye, Google AI}}.
We provided a web-based diary interface (Figure~\ref{fig:interface} in Appendix A) that participants could access whenever they felt the urge to use an LLM, asking questions to capture the context of and emotional responses to each moment. Participants were instructed to write a minimum of five diary entries per day.
Participants were also asked to complete a daily reflection via the same web interface at the end of each day to capture broader experiences and emotions throughout the day.
For detailed diary and reflection questions, see Appendix~\ref{appendix1} and Appendix~\ref{appendix2}, respectively.

\begin{table*}[t]
\centering
\caption{Participant demographics and LLM usage information. (For LLM-D12, we only utilized \textit{instrumental dependency} items.)}
\label{demo}
\small
\begin{tabular}{c c c c c c}
\toprule
ID & Age & Gender & Occupation & LLM Use/Day (avg.) & LLM-D12~\cite{yankouskaya2025llm} \scriptsize{(range: 6--36)} \\
\midrule
P1  & 25 & F & Graduate Student \scriptsize{(Computer Sci.)}        & 16--20 & 26 \\
P2  & 23 & F & DJ                                        & 21+    & 21 \\
P3  & 23 & F & Undergraduate Student \scriptsize{(Bio \& Brain Eng.)} & 21+    & 26 \\
P4  & 26 & F & Graduate Student \scriptsize{(Industrial Design)}      & 6--10  & 25 \\
P5  & 24 & F & Content Creator                           & 21+    & 16 \\
P6  & 26 & M & Junior Developer                                & 16--20 & 24 \\
P7  & 23 & M & Graduate Student \scriptsize{(Aerospace Eng.)}  & 21+    & 23 \\
P8  & 36 & F & Private English Instructor      & 6--10  & 26 \\
P9  & 26 & M & Graduate Student \scriptsize{(Electrical Eng.)} & 16--20 & 23 \\
P10 & 22 & M & Undergraduate Student \scriptsize{(Semiconductor Eng.)} & 11--15 & 30 \\
\bottomrule
\end{tabular}
\end{table*}
Following the four-day diary study, we conducted semi-structured interviews to gain a holistic understanding of participants’ experiences and reflections. Each interview was conducted either in person or online and lasted approximately one hour. Since we recognized that restricting participants from using LLMs in their work could potentially induce stress, we informed the participants in advance that they could quit the study at any time if they felt unable to continue. 
In addition, all personal information collected from participants was securely stored using the Firebase database. Each participant was assigned an anonymous id, and any identifying information was removed from the research dataset prior to analysis. Access to the encrypted data was restricted to administrators of the research team only. This research process and protocol was approved by the university’s Institutional Review Board (IRB), and all data handling and storage procedures complied with IRB and platform-level data protection guidelines.

\textbf{\textit{Data Analysis.}}
As a result of the study, we collected a total of 200 entries and 40 reflection logs, as well as 8 hours and 51 minutes of audio recordings that were fully transcribed prior to analysis.
We conducted a thematic analysis~\cite{reflect, theme} using the interview transcripts as the primary dataset,
while the diary entries and reflection logs provided additional contextual information to deepen our understanding of participants’ experiences. 
Following this stage, the three authors met to collaboratively develop a codebook. Each researcher closely examined the interview data to generate an initial set of codes. All authors then engaged in an iterative process of discussion and refinement, during which codes were compared, organized, merged, and revised in search of emergent themes and recurring patterns. This process continued until the research team reached consensus on the code structure and thematic interpretations. 

\section{Findings}

The goal of this study was to examine how pervasive LLMs have become embedded in everyday work practices by temporarily restricting their use. In this section, we report what the withdrawal of LLMs revealed, focusing on (i) workflow gaps revealed through LLM withdrawal, (ii) values participants reclaimed through self-directed work, and (iii) the perceived inevitability of LLM use in contemporary work environments.

\subsection{Workflow Gaps Revealed Through LLM Withdrawal
}
\label{3_1}


Being accustomed to LLM assistance, participants experienced the withdrawal period as uncomfortable and frustrating. 
Participants often compared LLM absence to the lack of everyday technologies that had become taken for granted in their routines to convey the extent of this discomfort.
For instance, P8 likened working without LLMs to living without a dishwasher or a robotic vacuum cleaner, while others compared it to lacking access to a convenience store (P1), a vehicle (P9), or familiar search tools like Google (P2, P6), as well as writing documents without tools such as Microsoft Word (P7).
As such, working without LLMs felt unfamiliar and uncomfortable, and this discomfort manifested in a range of gaps across participants’ everyday work practices.


\textit{\textbf{Asking People for Help Felt Socially Costly.}}
Participants who had grown used to seeking assistance from LLMs perceived asking other people for help as a social burden, assuming others would find it tiring and burdensome. Even when they required alternative perspectives, they preferred switching between LLM services, such as from ChatGPT to Grok, rather than consulting other people (P5). 
P10 explained that they avoided asking others because they feared being seen as a \textit{“finger prince/princess,”} a Korean slang term for someone who burdens others by asking easily searchable questions.
However, the withdrawal pushed participants to turn to human help, and some found it more helpful than LLMs (P2, P10). P9 noted how LLM use had hindered social interaction, saying \textit{``If this study (four days of LLM withdrawal) had gone on for longer, say, for a year, discussions between people would be more active.''}

\textit{\textbf{Processing Web Search Keywords Felt Excessive.}}
In the absence of LLM support, participants relied on conventional search engines such as Google and reported increased difficulty in retrieving information. They perceived the need to synthesize, reformulate, and iterate search keywords as excessive, noting that a process that had once been straightforward now felt unexpectedly difficult and cumbersome after becoming accustomed to LLM use. \textit{“ChatGPT gives you what you want in one go, but I had to put my mind into how I could get results with the least number of searches.”} (P10). P4 mentioned that it felt inefficient to translate their thoughts in ways a machine could understand, defining LLMs as \textit{``technology with added power to understand humans.''}

\textit{\textbf{Reading and Writing was Met with Lowered Standards.}}
Participants reported reduced motivation to comprehend texts or refine their writing when LLM assistance was unavailable. They viewed LLMs as tools that enable higher-quality performance. However, in their absence, participants were more willing to accept lower-quality outcomes rather than invest additional time and effort to improve their work, which they perceived as inefficient or wasteful. For example, P1 described a change in written communication habits, explaining, \textit{“When I write DMs or emails to my professor in English, I always ask ChatGPT for revision, but I decided I didn't have to write perfect English (during the study).”}



\textit{\textbf{Loss of Instant Assistance Prompted Delay or Avoidance.}}
Loss of immediate access to LLMs, which participants had grown accustomed to, led some to abandon certain tasks or wait until LLM use was available again. \textit{``I've been putting things off. I should start with them now that the study is over.''} (P1) The desire for immediate support lingered even when LLMs were withdrawn from their work, causing discomfort and frustration. P3 shared their experience of this absence, saying \textit{``I gave up tasks like organizing information that I usually would have asked an LLM to do. [...] Not being able to know what comes into my mind straight away was a bigger thing than I expected."} (P3)

\textit{\textbf{Difficulties Adapting to the Change in Work Pace.}} 
Participants often underestimated how long tasks would take without LLM assistance, showing that they were not fully aware of how much LLMs had sped up their work. As they struggled to adjust to this slower pace of work, participants responded in different ways, including frustration, anxiety, and changes to their work routines. For example, P2 expressed frustration with the perceived inefficiency, stating, \textit{“I had to look up information on some equipments, but it took so long I didn't think it was worth the time.”} Similarly, P8 described feeling nervous upon realizing how much longer it took to find materials without LLM support, which led them to adjust their schedules or work overtime.

\subsection{Work Values are Reclaimed by Taking Back LLM-Delegated Tasks
}
Despite immediate difficulties in adapting to workflows without LLMs, as discussed in Section \ref{3_1}, participants also recognized the value of working independently. We report three work-related values: (i) clarity in work through constructing and tracing their own reasoning, (ii) an increased sense of ownership over outcomes, and (iii) heightened awareness of personal priorities between tasks.



\textit{\textbf{Clarity in Work.}}
Without LLMs in their workflow, participants reported a clearer understanding of their work because they had to construct and trace the underlying logic themselves. 
In contrast, when using LLMs, participants described a tendency to accept the output without fully unpacking each step. The model often gave reasoning that felt misaligned with their original intent, even when participants attempted to redirect or correct the output: \textit{``I try to follow the LLM's logic, but I get frustrated because I have to keep telling it no, that's not what I meant.''} (P3)
Engaging with information in greater depth and from a wider range of sources without LLM assistance enabled participants to achieve greater clarity, which also contributed a better sustained understanding of the material.

\textit{\textbf{Increased Sense of Ownership.}} 
When participants completed tasks without LLMs, they reported a stronger sense of ownership over the outcomes. They described work produced with LLM assistance as feeling \textit{“not their own”} (P1, P3, P7), since less time and effort was invested in refining the work themselves.
By contrast, completing the same tasks on their own–\textit{“not through some other technology (i.e., LLMs)”} (P5)–required participants to engage in the entire process from ideation to decision-making resulting in outputs that were done entirely on their own which fostered a stronger sense of ownership and pride: \textit{“I had to do some ideation, this time without LLM, and it really felt like it was mine.”} (P1).



\textit{\textbf{Awareness of Personal Priorities Between Tasks.}}
As completing tasks without LLM assistance required more time and effort, participants had to be more selective about which tasks to prioritize, as they could handle fewer tasks than when LLM support was available. This constraint prompted participants to re-evaluate where their time and effort should be best spent, revealing that they had previously delegated even core, profession-defining tasks to LLMs for efficiency. Through this process, participants (P1, P5, P6, P7; graduate student, content creator, private english instructor, junior developer) regained clarity about which skills and activities were crucial to their profession. 
Reflecting this shift in prioritization, P1 explained, 
\textit{“For me, it’s a matter of priority. Since I’m a reseacher, not a developer, [...] I don’t think I’ll ever be in a situation where I have to write code entirely from my own head without any LLM support. I don't think learning to code is something I need to invest in.”}



\subsection{Inescapable Dependency on LLMs}

The withdrawal of LLMs enabled participants to recognize their dependency in their work environment. Working without LLMs was difficult not only because individuals were too dependent to work without LLMs, but also because of a social environment that viewed LLM use as normative.

\textit{\textbf{Too Dependent to Work Without LLMs.}}
All participants reported that the use of LLMs impeded skill development and acquisition. Many described LLMs as no longer simply assisting their work but replacing core aspects of it, leading them to feel unable to work without LLM support. For instance, P6 described themself as \textit{“a developer who can’t even catch a bug”}, attributing this to the habitual copying and pasting between their code and LLM outputs without fully understanding the surrounding context.

This dependency also shaped participants’ motivation and attitudes toward learning. Some participants avoided tasks they had only ever completed with LLM assistance, expressing little willingness to learn how to do them independently. P3 explained that they would not have participated in the study had they been required to do a coding assignment during that period, stating, \textit{“I've never done it (the coding task) without ChatGPT. I don't think I can do it at all without it. [...] But I don’t think I really have the will to learn it either. ChatGPT helps me so well anyway.”} 
P8 compared this growing reliance to a gradual loss of initiative: \textit{“It’s like how when you’re standing up you want to sit down, and when you’re sitting you want to lie down, and when you’re lying down, you want to sleep. I could have just sat there, but it’s like the LLM is trying to send me to sleep.”}

\textit{\textbf{Socially Inevitable Use of LLMs.}}
Participants regarded the use of LLM as inevitable, considering it not as an optional tool but as a requirement for remaining competitive at work. They described the inability to use LLMs during the study as a disadvantage, as LLM use was seen as part of their basic work capacity rather than a privilege. For instance, P3 commented \textit{``everyone is using ChatGPT anyway.''} P9 saw not using LLMs as a personal loss, saying \textit{``I used to think using LLMs was cheating, but now I think you're the only one losing out if you don't.''}

Participants further perceived LLM use as not just a personal choice but as a new social norm pressuring those who do not adopt it. P1 explained there is no such thing as excessive when it comes to using LLMs, saying \textit{``I used to think, why would you ask LLM something so trivial, but now LLMs are everywhere. There is no line between what you can ask for help and what you can't.''} P7 elaborated on the social pressure experienced within their research lab: \textit{``Now even my professor tells me to just get it (paperwork) over with with an LLM, and I feel it can't be helped in this era.''}

\section{Discussions}


LLM withdrawal reveals that LLMs have already become deeply embedded in everyday work practices, functioning not merely as optional tools but as an implicit work infrastructure.
Participants’ common discomfort, frustrations, and task delays or avoidances during withdrawal indicate that many current workflows are already optimized around the expectation of instant LLM assistance that is taken for granted. 
From this perspective, the workflow gaps surfaced through LLM absence should not be regarded as individual shortcomings or skill degradation, but rather as traces of technology adaptation in which work practices have been reconfigured.
While previous research on LLM dependency focused on the effects of the dependency on cognitive efforts and how it can be measured on an individual level~\cite{kasneci2023chatgpt, peterson2025ai, lee2025impact, lo2024influence, jia2025effect, yankouskaya2025llm, li2025assistance}, we conceptualize this dependency from an infrastructural role of LLMs. This, in turn, normalizes their use in everyday work and generates social pressure to adopt them.
Together, these findings suggest that the central challenge is no longer simply whether to use LLMs or not, but how their use should be negotiated, for which tasks, and to what extent. 


In this light, our study highlights an opportunity for the HCI community to design work practices that move beyond the binary framing of use versus non-use.
Withdrawal made work values more visible that had been obscured by habitual delegation to LLMs, including clarity, ownership, and professional integrity. While these values do not align with those assumed in the design of LLMs such as generativity~\cite{sun-etal-2023-text}, productivity~\cite{productivity_ChatGPT}, or proactivity~\cite{Diebel2025ProactiveAI}, they are crucial to sustain workers' professional identity.
Building on the concept of technological appropriation \cite{DourishAppropriation}, the process by which people adopt technologies not just as intended by their providers but in ways that better fit their work practices and contexts, we envision \textbf{value-driven appropriation} as a mode of LLM use. Here, workers deliberately shape LLM engagement in ways that align with their professional values rather than those pre-defined by LLMs.
One possible scenario along this line includes prior work in education~\cite{mollick2023student, park2024chatgpt_writing_feedback}, which examined cases of LLM use where structured prompting is used as a strategy to support student agency by restricting LLMs' response until students' intent is sufficiently articulated.
Design practices on LLM use could create room for users to negotiate and sustain their own professional standards.




\section{Limitations, Future Work, and Conclusion}
Our study has limitations in that all participants were South Korean, a country with the second-largest number of paid ChatGPT users in the world \cite{OpenAI_Korea_2025}. Our focus within this population was on those with frequent use of and high dependency on LLMs. Participants’ cultural and national contexts may play an important role in shaping their perceptions and use of LLMs. Therefore, similarly designed studies conducted in different cultural contexts may yield different findings, highlighting the need for future work with more diverse participant populations. Regarding the diary study period, although a four-day study was sufficient to generate meaningful findings in our context, a longer-term study (e.g., over a month) may yield additional or different insights. In addition, future work could investigate study designs that encourage participants to explicitly plan and reflect on their LLM use strategies in everyday work. Observing how individuals deliberately configure how, and for which tasks they engage with LLMs would offer deeper insight for designing personalized and adaptive forms of LLM appropriation in practice.




To conclude, we conducted a four-day diary study in which participants were asked to refrain from using LLMs. This withdrawal revealed the difficulties that emerged from working in LLM-accustomed environments, values participants rediscovered in working independently, and the inescapable ways in which LLMs had become embedded in everyday work practices. 
Our study offers insights for designing LLM-assisted work practices that recognize the infrastructural role of LLMs while intentionally considering human collaboration and professional agency.

\begin{acks}
This work was supported by the National Research Foundation of Korea (NRF) grant funded by the Korea government (MSIT) (No. RS-2021-NR059056).
\end{acks}

\bibliographystyle{ACM-Reference-Format}
\bibliography{sample-base}

\appendix
\section{Study Details}
\label{appendix}

\begin{figure}
    \centering
    \includegraphics[width=\linewidth]{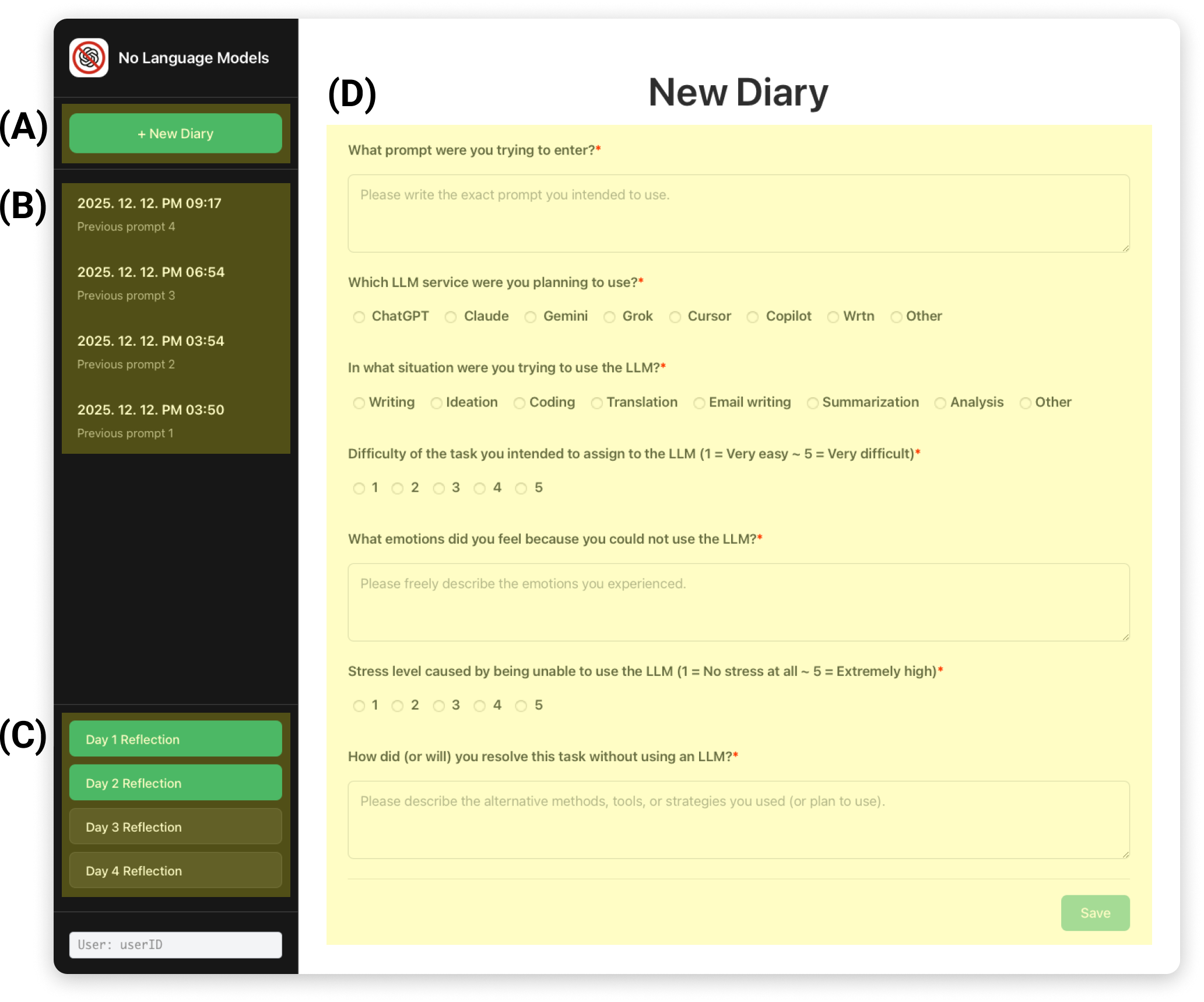}

    \caption{Web-based interface used for the diary study. (This is a translated version of the original diary study.) (A) New diary entry button. (B) Previous diary list. (C) Daily reflection button. (D) Diary entry and reflection form.}
    \Description{Screenshot of the diary interface used during the diary study.}
    \label{fig:interface}

\end{figure}

\subsection{Diary Study Questions}
\label{appendix1}

\begin{itemize}
    \item What prompt were you trying to enter? (short answer question)
    \item Which LLM service were you planning to use? (multiple choice question)
    \item In what situation were you trying to use the LLM? (multiple choice question)
    \item Difficulty of the task you intended to assign to the LLM (1 = very easy, 5 = very difficult).
    \item What emotions did you feel because you could not use the LLM? (short answer question)
    \item Stress level caused by being unable to use the LLM (1 = no stress at all, 5 = extremely high).
    \item How did (or will) you resolve this task without using an LLM? (short answer question)
\end{itemize}

\subsection{Daily Reflection Questions}

\label{appendix2}

\begin{itemize}
    \item At what moment today did you most strongly want to use an LLM?
    \item How did you feel after completing your work without LLM assistance?
    \item How did you complete your work without the help of an LLM?
    \item Were there any tasks that you could not complete without LLM assistance?
    \item How efficient was your work without LLM assistance?
    \item Compared to when you use an LLM, how satisfied were you with your work today?
    \item Please describe any positive changes or benefits you experienced while working without LLM assistance.
    \item Please describe any difficulties or drawbacks you experienced while working without LLM assistance.
\end{itemize}

\subsection{Semi-Structured Interview Questions}

\begin{itemize}
    \item Please describe your overall experience of living and working without LLM assistance during the four-day period.

    \item Which task did you want to use an LLM for during the withdrawal?

    \item What was the best or worst moment of working without an LLM?
    \item What (dis)advantages did you notice from not using an LLM?

    \item Did you rediscover or relearn any skills or knowledge during this period?

    \item Were there any unexpected encounters with LLMs?

    \item What was the task that felt most trivial that you wanted to delegate to LLMs?
    
    \item Was there a moment when you realized that you could handle a task on your own more easily than you had expected?

    \item What tasks do you feel comfortable or guilty when delegating to LLMs? What do you think is the cause of this difference?


    \item If life without LLMs continued, how do you think it would affect your work?

    \item Did participating in this study change your perception of LLMs? If so, how might this influence your future use of LLMs?

    \item If you used an LLM during the withdrawal period, please describe with honesty.

\end{itemize}

\end{document}